\documentclass[12pt]{article}

\pdfoutput=1

\usepackage{cite}

\usepackage{amssymb,latexsym}

\usepackage{epstopdf}

\usepackage{graphics,epsfig}

\usepackage{color}

\usepackage{mathrsfs}

\DeclareMathAlphabet{\mathpzc}{OT1}{pzc}{m}{it}

\let\a=\alpha \let\b=\beta \let\g=\gamma \let\d=\delta \let\e=\epsilon
  \let\th=\theta  \let\k=\kappa
\let\l=\lambda \let\m=\mu \let\n=\nu \let\x=\xi \let\p=\pi %\let\r=\rho
\let\s=\sigma \let\t=\tau  \let\f=\phi  
 
        \let\Th=\Theta 
\let\X=\Xi  \let\S=\Sigma  \let\Y=\Psi
 
\let\la=\label  
  
 \def\bd{\begin{document}} \def\ed{\end{document}}
\def\ds{\documentstyle} \let\fr=\frac \let\bl=\bigl \let\br=\bigr
\let\Br=\Bigr \let\Bl=\Bigl
\let\bm=\bibitem
\let\na=\nabla
\def\tU{{\widetilde U}}
\let\pa=\partial \let\ov=\overline
\def\ie{{\it i.e.\ }}
\newcommand{\be}{\begin{equation}}
\newcommand{\ee}{\end{equation}}
\def\ba{\begin{array}}
\def\ea{\end{array}}
\def\ft#1#2{{\textstyle{{\scriptstyle #1}\over {\scriptstyle #2}}}}
\def\fft#1#2{{#1 \over #2}}
\def\F#1#2{{ F_{#1}^{(#2)} }}
\def\cF#1#2{{ {\cal F}_{#1}^{(#2)} }}

\def\R{{\bf R}}
\def\sst#1{{\scriptscriptstyle #1}}
\def\oneone{\rlap 1\mkern4mu{\rm l}}
\def\e7{E_{7(+7)}}
\def\td{\tilde}
\def\wtd{\widetilde}
\def\im{{\rm i}}
\def\bog{Bogomol'nyi\ }
\newcommand{\ho}[1]{$\, ^{#1}$}
\newcommand{\hoch}[1]{$\, ^{#1}$}
\newcommand{\bea}{\begin{eqnarray}}
\newcommand{\eea}{\end{eqnarray}}
\newcommand{\ra}{\rightarrow}
\newcommand{\lra}{\longrightarrow}
\newcommand{\Lra}{\Leftrightarrow}
\newcommand{\ap}{\alpha^\prime}
\newcommand{\bp}{\tilde \beta^\prime}
\newcommand{\cB}{{\cal B}}
\newcommand{\cO}{{\cal O}}
\newcommand{\vecx}{\vec{x}}
\newcommand{\vecy}{\vec{y}}
\newcommand{\vecp}{\vec{p}}
\newcommand{\vecq}{\vec{q}}
\newcommand{\tr}{{\rm tr} }
\newcommand{\Tr}{{\rm Tr} }
\newcommand{\NP}{Nucl. Phys. }

\newcommand{\cL}{{\cal L}}
\newcommand{\cA}{{\cal A}}
\newcommand{\cT}{{\cal T}}
\newcommand{\cD}{{\cal D}}
\newcommand{\cH}{{\cal H}}

\def\sst#1{{\scriptscriptstyle #1}}
\def\0{{\sst{(0)}}}
\def\1{{\sst{(1)}}}
\def\2{{\sst{(2)}}}
\def\3{{\sst{(3)}}}
\def\4{{\sst{(4)}}}
\def\5{{\sst{(5)}}}
\def\6{{\sst{(6)}}}
\def\7{{\sst{(7)}}}
\def\8{{\sst{(8)}}}
\def\9{{\sst{(9)}}}
\def\p{{\sst{(p)}}}
\def\q{{\sst{(q)}}}
\def\ve{\varepsilon}
\def\vf{\varphi}
\def\F{\Phi}
\def\wg{\wedge}

\def\thb{\bar{\theta}}
\def\Thb{\bar{\Theta}}
\def\barp{\bar{p}}
\def\barq{\bar{q}}
\def\barc{\bar{c}}
\def\bard{\bar{d}}
\def\e{\epsilon}

\def \bi{\bibitem}
\def \la {\label}

\def \l {\lambda}
\def\foot{\footnote}
\def \tl  {{\tilde \l}}
\def \sql {{\sqrt \l}}
\def \adss {$AdS_5 \times S^5$\ }
\newcommand{\rf}[1]{(\ref{#1})}
\def \ov {\over}

\def\th{\theta}
\def\Th{\Theta}
\def\vth{\vartheta}
\def\btheta{{\bar\theta}}
\def\ttheta{{{\tilde\theta}}}
\def\bttheta{{{\bar\ttheta}}}
\def\vth{\vartheta}

\def\ra{\rightarrow}
\def\N{\nabla}
\def\F{{\cal F}}
\def\uM{\underline{M}}
\def\uA{\underline{A}}
\def\uN{\underline{N}}
\def\uP{\underline{P}}
\def\ua{\underline{a}}
\def\ub{\underline{b}}
\def\uc{\underline{c}}
\def\ud{\underline{d}}
\def\ue{\underline{e}}
\def\uf{\underline{f}}
\def\ui{\underline{i}}
\def\uj{\underline{j}}
\def\uk{\underline{k}}
\def\ul{\underline{l}}
\def\ual{\underline{\alpha}}
\def\ube{\underline{\beta}}
\def\um{\underline{m}}
\def\un{\underline{n}}
\def\up{\underline{p}}
\def\uq{\underline{q}}
\def\ur{\underline{r}}
\def\us{\underline{s}}
\def\umu{\underline{\mu}}
\def\unu{\underline{\nu}}
\def\ula{\underline{\l}}
\def\uka{\underline{\k}}
\def\usi{\underline{\s}}
\def\urh{\underline{\r}}
\def\cc{\circ}
\def\eqv{\equiv}

\def\ni{\noindent}

\def\Ep{E^{{}^{(+)}}}
\def\Em{E^{{}^{(-)}}}

\def\Mp{M^{{}^{(+)}}}
\def\Mm{M^{{}^{(-)}}}

\def \ha{{1\ov 2}}

\def\r{\rho}

\def\Y{{\rm Y}}
\def\X{{\rm X}}
\def\tY{\tilde{\rm Y}}
\def\tX{\tilde{\rm X}}
\def\dY{\dot{\rm Y}}
\def\dX{\dot{\rm X}}

\def \J {\mathcal{J}}
\def \del {\partial}

\def\dF{\dot{F}}
\def\dG{\dot{G}}
\def\df{\dot{f}}
\def \E {{\cal E}}
\def \S {{\cal S}}
\def \J {{\cal J}}

\def\ms{\mathcal{S}}
\def\mj{\mathcal{J}}
\def\soj{\fr{\ms}{\mj}}
\def \R {{\bf R}}
\def \om {\omega}
\def \bE {\bar E}
\def \x {{\cal X}}

\def \bi{\bibitem}
\def \la {\label}

\def \l {\lambda}
\def\foot{\footnote}
\def \tl  {{\tilde \l}}
\def \sql {{\sqrt \l}}
\def \adss {$AdS_5 \times S^5$\ }
\def \ov {\over}

\def \varpi {{\rm w}}

\def\thb{\bar{\theta}}
\def\Thb{\bar{\Theta}}
\def\mb{\bar{\m}}
\def\ab{\bar{\a}}
\def\zb{\bar{z}}
\def\psib{\bar{\psi}}
\def\barp{\bar{p}}
\def\barq{\bar{q}}
\def\barc{\bar{c}}
\def\bard{\bar{d}}
\def\e{\epsilon}
\def\wb{\bar{w}}
\def\lb{\bar{\l}}
\def\Jb{\bar{J}}
\def\Nb{\bar{N}}
\def\Zb{\bar{Z}}
\def\pab{\bar{\pa}}

\def\At{\tilde{A}}
\def\Bt{\tilde{B}}
\def\Ct{\tilde{C}}
\def\Dt{\tilde{D}}
\def\Et{\tilde{E}}
\def\Ft{\tilde{F}}
\def\Gt{\tilde{G}}
\def\Ht{\tilde{H}}
\def\Kt{\tilde{K}}
\def\Mt{\tilde{M}}
\def\Nt{\tilde{N}}
\def\Rt{\tilde{R}}
\def\at{\tilde{a}}
\def\bt{\tilde{b}}
\def\ct{\tilde{c}}
\def\dt{\tilde{d}}
\def\et{\tilde{e}}
\def\ft{\tilde{f}}
\def\htil{\tilde{h}}
\def\gt{\tilde{g}}
\def\nt{\tilde{n}}
\def\mut{\tilde{\mu}}
\def\nut{\tilde{\nu}}
\def\pht{\tilde{\f}}
\def\vft{\tilde{\vf}}

\def\rht{\tilde{\rho}}

\def\asth{\hat{*}}
\def\phh{\hat{\phi}}

\def\bA{{\bf A}}

\def\ola{\overleftarrow}
\def\ora{\overrightarrow}
\def\alt{\tilde{\a}}

\def\eh{\hat{e}}
\def\eph{\hat{\e}}
\def\ph{\hat{p}}
\def\alh{\hat{\a}}
\def\beh{\hat{\b}}
\def\gah{\hat{\g}}
\def\Fh{\hat{F}}
\def\muh{\hat{\m}}
\def\nuh{\hat{\n}}
\def\thh{\hat{\th}}
\def\rhh{\hat{\r}}
\def\dh{\hat{d}}
\def\ih{\hat{i}}
\def\jh{\hat{j}}
\def\hh{\hat{h}}
\def\nh{\hat{n}}
\def\gh{\hat{g}}
\def\kh{\hat{k}}
\def\deh{\hat{\d}}
\def\wh{\hat{w}}
\def\lah{\hat{\l}}
\def\Ah{\hat{A}}
\def\Kh{\hat{K}}
\def\Nh{\hat{N}}
\def\Rh{\hat{R}}
\def\Ch{\hat{C}}
\def\Omh{\hat{\Omega}}

\def\xh{\hat{x}}

\def\ps{\rlap{\, /}\;\,p }
\def\ks{\rlap{\, /}\;\,k }

\def\gym{g_{YM}}

\def\adot{\dot{a}}
\def\bdot{\dot{b}}
\def\bpa{\bar{\pa}}

\def\pr{\prime}
\def\ssk{\medskip}
\def\clb{\color{blue}}
\def\clr{\color{red}}
\def\clg{\color{green}}

\def\bfA{{\bf A}}
\def\bfB{{\bf B}}
\def\bfK{{\bf K}}
\def\bfU{{\bf U}}
\def\bfX{{\bf X}}
\def\bfY{{\bf Y}}
\def\bfg{{\bf g}}
\def\bfn{{\bf n}}

\begin{document}

\overfullrule=0pt
\parskip=2pt
\parindent=12pt
\headheight=0in \headsep=0in \topmargin=0in
\oddsidemargin=0in

\vspace{ -3cm}
\thispagestyle{empty}
%\vspace{1cm}
%\begin{flushright}
%Preprint DFPD 01/TH/\\
%hep-th/
%\end{flushright}

 \vspace{0.1cm}

\setcounter{equation}{0}
\setcounter{footnote}{0}
\setcounter{section}{0}

\begin{center}

{\Large\bf Mathematical foundation of foliation-based quantization}

\vskip 0.8cm

 \vspace{.5cm}

\vspace{0.5cm}
I. Y. Park
\\

\vspace{0.3cm}

%{\it Center for Quantum Spacetime, Sogang University\\
%Shinsu-dong 1, Mapo-gu, 121-742 South Korea \\
%}

{\it Department of Physics, Hanyang University \\
Seoul 133-791, Korea}\\

\vspace{0.3cm}
{\it Department of Applied Mathematics,
Philander Smith College %\footnote{Home institute}
                               \\
Little Rock, AR 72223, USA \\
inyongpark05@gmail.com
}

\end{center}

 \vspace{0.1cm}

 \begin{abstract}
 %%%%%%%%%%%%%%%%%%%%%%%%%%%%%%%%%

\ni We have recently proposed in \cite{Park:2014tia} the quantization of pure 4D Einstein gravity through hypersurface foliation, and observed that the 4D Einstein gravity becomes renormalizable once all (or most) of the unphysical degrees of freedom are removed. In this work, we confirm this observation from a more mathematical angle. In particular, we show that the physical state condition arising from the shift vector constraint connects with the requirement that the manifold admit ``totally geodesic (TG) foliation". The TG foliation, in turn, makes it possible to view the 4D manifold as abelian fibration over a 3D base. Associating the abelian fibration with the 4D diffeomorphism leads to reduction of the 4D manifold to 3D, thereby realizing and generalizing the holography of 't Hooft.  
\end{abstract}
\newpage
%%%%%%%%%%%%%%%%%%%%%%%%%%%%%%%%%%%%%
%%%%%%%%%%%%%%%%%%%%%%%%%%%%%%%%%%%%%
\section{Introduction}
%%%%%%%%%%%%%%%%%%%%%%%%%%%%%%%%%%%%%
%%%%%%%%%%%%%%%%%%%%%%%%%%%%%%%%%%%%%
We have recently proposed in \cite{Park:2014tia} the quantization of pure 4D Einstein gravity through hypersurface foliation, and observed that the 4D Einstein gravity becomes renormalizable once all (or most) of the unphysical degrees of freedom are removed.\footnote{The approach of \cite{Park:2014tia} and the present work has now been more fully developed in \cite{Park:2015qxa,Park:2016zgt}. (See the references therein as well.) The most important thing achieved by the full gauge-fixing is the determination of the 3D physical states; one may adopt a more 3D-oriented route or 4D-oriented method depending on one's purpose \cite{Park:2014noa}.} $\!$The main motivation for the present work is finding the origin of this observation by employing (more) concrete and quantitative mathematical foundations of foliation theory. 
In doing so, additional insights into the connection with 't Hooft's holography are gained.

The interplay between a bulk spacetime and its boundary has been the central theme for recent progress in theoretical physics. We employ the mathematics of foliation theory at a quantitative level in a continued effort to extend the theme to the interplay between a bulk and its hypersurfaces. 
Intuitively speaking, foliation is a way of viewing a manifold as constructed by putting  together lower-dimensional ``similar"-shaped manifolds.    
The notion of foliation led to some interesting results \cite{Sato:2002kv}\cite{Hatefi:2012bp} in the context of AdS/CFT. More recently, it has been employed in \cite{Park:2013iqa,Park:2013bma, Park:2014mba} to uncover various aspects of black hole information. 
It is highly likely that the notion will lead to significant and as yet unprecedented results in gravitational physics.

A more mathematically precise look at the possibility explored in this work will be found in the main body; a rough overview of this possibility is as follows. Suppose a 4D manifold is constructed by an abelian group fibration over a 3D base manifold. If the group action can be associated with the gauge symmetry, the physical Hilbert space would be reduced to that of the base manifold. We show below that the gravity holography as observed in \cite{Park:2014tia} is exactly of this type.

We focus on a class of spacetimes called the globally hyperbolic spacetimes. (They cover most of the cosmologically interesting spacetimes.)
A globally hyperbolic spacetime admits a codimension-1 foliation through a family of hypersurfaces. 
In general, it is not guaranteed that a globally hyperbolic spacetime will admit an alternative construction as a principal bundle with an abelian structure group, a setup used for possible holographic reduction in the paragraph above.
The key issue then is to establish the possible connection between these two dual views: the manifold viewed as codimension-1 foliation and as a principle bundle of 1D fibration over the 3D base. We observe that the condition obtained in \cite{Park:2014tia} that arose from the shift vector constraint is what connects these two views.

As a matter of fact, the two views are related by a ``duality". The duality involved is a mathematical one and operates between two different foliations: the Riemannian and totally geodesic. The shift vector constraint makes a direct connection with the Riemannian foliation. Happily, the totally geodesic foliation is the precise form of the 1D abelian fibration mentioned above.   
%%%%
\begin{figure}
\centerline{
\begin{minipage}[b]{8cm}
             \epsfxsize=8cm
              \epsfbox{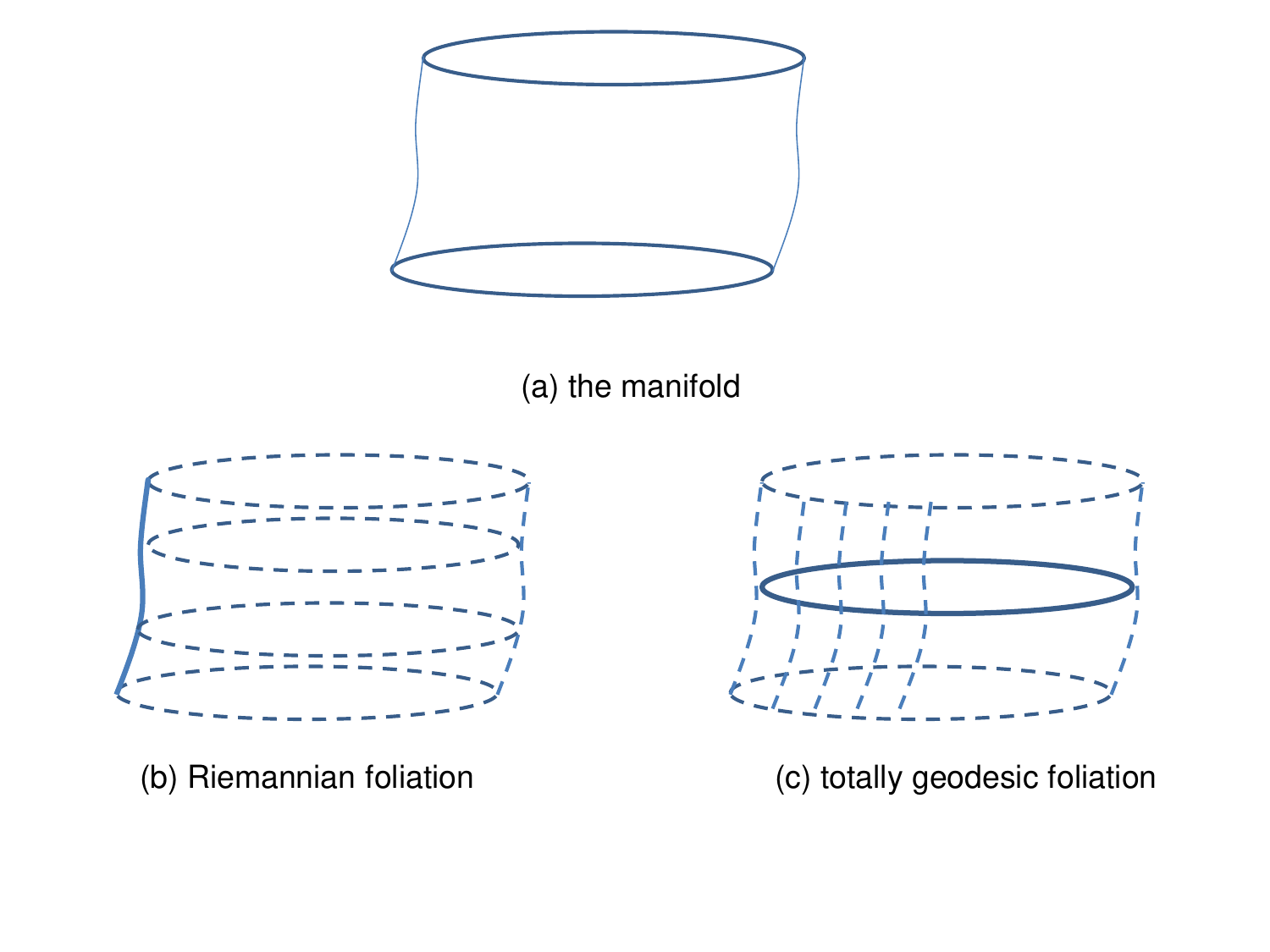}
      \end{minipage}
      }
\caption{duality in foliations }
\label{fig}
\end{figure}
 
\vspace{.3in}
\ni The rest of the paper is organized as follows.
In the next section, we review necessary elements of differential geometry and lay out mathematical foundations to confirm the observation made in \cite{Park:2014tia}. The duality between totally geodesic foliation and Riemannian foliation will be one of the central ingredients for establishing the gravity holography. In section 3, we start by reviewing the analysis of the shift vector constraint in \cite{Park:2014tia}.\footnote{There is another constraint arising from fixing the lapse function, the analogue of the Hamiltonian constraint of the Hamiltonian formulation. We focus on the shift vector constraint in this work. } We point out that the condition from the shift vector constraint, \rf{Riemannconp}, makes the foliation `Riemannian' in the jargon of foliation theory. The Riemannian foliation under consideration admits a dual foliation called the totally geodesic foliation.\footnote{It appears that this is not true in general: an additional condition on the integrability of the orthogonal foliation should be satisfied. } A manifold of totally geodesic foliation has associated Lie algebra and the dimension of the Lie algebra is one in the present case. As the second key step, we identify this abelian group as the origin of the gauge symmetry associated with the 4D metric. 
The gauge-fixing then corresponds to taking the quotient of the bundle by the group, thereby, realizing the holography. We conclude with future directions in the final section.

%%%%%%%%%%%%%%%%%%%%%%%%%%%%%%%%%%%%%
%%%%%%%%%%%%%%%%%%%%%%%%%%%%%%%%%%%%%
\section{Review of differential geometry}
%%%%%%%%%%%%%%%%%%%%%%%%%%%%%%%%%%%%%
%%%%%%%%%%%%%%%%%%%%%%%%%%%%%%%%%%%%%
In this section, we review and introduce elements of differential geometry with the goal of placing the observation in \cite{Park:2014tia} on concrete mathematical ground. The goal requires advanced-level differential geometry. We review intermediate-level differential geometry (in particular, the coordinate-free formulation) for notational uniformity, and refer to mathematical literature for several key mathematical theorems.    
There are several graduate-level textbooks and reviews on general relativity, such as \cite{Wald,Poisson,Gron,Gourgoulhon:2007ue}, that employ intermediate-level differential geometry. The present work requires some crucial results such as eq.\rf{liecocom}  below from advanced-level differential geometry discussed, e.g., in \cite{Kobayashi1,Kobayashi2,Nakahara,Moerdijk,Gromoll}. Those results can only be derived in a setup wherein the covariant derivative and Lie derivative are defined through a one-parameter family of transformations.
Properly covering the needed elements of differential geometry would put this section out of proportion; we present minimized materials, intending it as a guide to more detailed, in-depth accounts in the literature.

\vspace{.2in}

The so-called ``coordinate-free" formulation has been widely used in modern differential geometry. (Indeed we will use some of the results obtained in that formulation in the main section). Less familiar though it may be to physicists, the coordinate-free notation has proven useful and powerful: it has provided simpler proofs of existing theorems, and new mathematical insights have been gained. It has also led to many new results that would have been harder to obtain with the conventional index notation.

The starting point of the coordinate-free notation is a coordinate-free definition of a vector (and a tensor). There exists an isomorphism between the conventional and new definitions of a vector. The definition of a vector most familiar to physicists, i.e., the conventional definition, is an $n$-component entry of numbers that transforms according to fixed rules under the coordinate change dictated by the symmetry group of the system. In the coordinate-free formulation, a vector is defined as the tangent of a curve on the manifold. This definition, in turn, introduces an equivalence class by criterion of having the same tangent. A vector can also be viewed as a differential operator that acts on the function space of the manifold: a vector is like a set of partial derivatives.

Still another way to view a vector is based on the directional derivative.
Consider a manifold $M$ and a curve $x(t)$ with a fixed range of the parameter $t$. A vector $X$ at $x(t)\in M$ (a particular value of $t$ being considered) can be defined as a map from the function space to a real number given by
%%%
\bea
\bfX f=\fr{df(x(t))}{dt}
\eea
%%%
The collection of  vectors $\bfX$ forms the tangent space, $T(M)$, at $x=x(t)$.
Two more types of derivatives - which commute with contraction\footnote{Contraction is the coordinate-free version of the index contraction.} - are essential on a curved manifold. The first is the Lie derivative, and is associated with infinitesimal coordinate transformation (the general coordinate transformation in physics). In mathematics, it is defined through a finite group action, $\vf_t$, called a one-parameter group of transformations that satisfy
%%%
\bea
\vf_{s+t}(x)=\vf_{t}(\vf_s(x))
\eea
%%% 
where $x$ represents a point in $M$, $x\in M$.
The Lie derivative of a tensor $\bfK$ along a vector $\bfX$ is defined by
%%%
\bea
\mathscr{L}_\bfX  \bfK=\lim_{t\ra 0}\fr{\bfK'_{\vf(x)}-\bfK_{x} }{t}
\la{liedef}
\eea
%%%
where the $'$ represents how the field $\bfK$ transforms under the group. The 
second derivative is the covariant derivative, and it requires more structure. 
We turn to a principle fiber bundle before we define the covariant derivative.

A (differentiable) principle fiber bundle of dimension $n$, denoted by $M(B,G,\pi)$, is a manifold with the action of the group $G$ defined by
%%%
\bea
R_a x:(x,a) \ra xa 
\eea
%%% 
where $(x,a) \in M\times G$. The letter $R$ stands for right multiplication; alternatively, one can use left multiplication. The bundle is also equipped with the projection 
map $\pi$ such that $\pi(x)=y$ with $y$ being a point on the base manifold $B$. The base manifold $B$ is the quotient space $B=M/G$. The group action on a point of the base manifold $B$ ``generates" the fiber and every fiber is diffeomorphic to $G$. The tangent space $T(M)$ can be decomposed into a horizontal component and a vertical component, a notion central to the covariant derivative:
%%%
\bea
 \bfX &=& \bfX^h+\bfX^v,\quad \bfX^h \in {\cal H}\;,\; \bfX^v \in {\cal V};
\eea
%%%
%%%
\bea
T(M)&=& {\cal H} \oplus {\cal V} 
\eea
%%%
where ${\cal H}$ (resp. ${\cal V}$) represents the horizontal subspace ${\cal H}$ (resp. vertical subspace)
The vertical subspace ${\cal V}$ consists of vectors tangent to the {\em fiber} through $x$ and ${\cal H}$ is its orthogonal complement in $T(M)$.

To define the covariant derivative, let us first define a horizontal lift of a curve and parallel displacement of fibers.\footnote{The definition of covariant derivative through 
parallel displacement of fibers eventually leads to the more familiar component definition through Christoffel symbols. However, we will later have a crucial use for the coordinate-free definition given here.} Let $\t\equiv x_t,\; 0\leq t \leq 1$ be a curve on the base manifold $B$ and $x_0$ be a point on the bundle $M$ such that $\pi(x_0)=y_0$. A horizontal lift (or simply a lift) of $\t$ is a curve whose tangent vectors are horizontal. 
There exists a unique horizontal lift, denoted by $\t^*$, of $\t$ through $x_0$; its endpoint $x_1$ maps to $y_1$ via the projection: $\pi(x_1)=y_1$. Varying $x_0$ along the fiber $\pi^{-1}(y_0)$, one gets a mapping - which can be shown to be an isomorphism - between the two fibers $\pi^{-1}(y_0)$ and $\pi^{-1}(y_1)$. This mapping is called the parallel displacement of the fibers, and will be denoted by the same letter $\t$ by following the convention in the mathematical literature.

With this we can define the covariant derivative of a section, $\vf$ of the bundle.\footnote{Strictly speaking, the section is a section of an associated bundle. (The definition of the associated bundle of a principle bundle can be found, e.g., in \cite{Kobayashi1} or \cite{Nakahara}.)} Given a curve $\t\equiv x_t$ and its tangent vector $\dot{x}_t$ (the dot denotes the time derivative), the covariant derivative $\N_{\dot{x}_t} \vf$ is defined by
%%%
\bea
\N_{\dot{x}_t} \vf=\lim_{\d t\ra 0}\fr{\t_t^{t+\d t}(\vf(x_{t+\d t}))-\vf(x_t)}{\d t}
\la{covaridef}     
\eea
%%%
where $\t_t^{t+\d t}$ denotes the parallel displacement from the fiber $\pi^{-1}(x_{t+\d t})$ to $\pi^{-1}(x_{t})$. By using the definitions of the Lie derivative \rf{liedef} and the covariant derivative \rf{covaridef},
one can show the following relation:
%%%
\bea
[\mathscr{L}_{\bfX}, \N_{\bfY}]=\N_{[\bfX,\bfY]} \la{liecocom}
\eea
%%%
 A proof of this relation can be found in chapter VI of \cite{Kobayashi1}. This relation was used in a crucial way in \cite{Park:2014tia} as we will review below.

The mathematical setup that we need in section \ref{main}, the main section, is categorized as ``Riemannian" or ``metric" foliation in mathematical literature.
Foliation can be viewed as a generalization of fibration. Only fibers of the same topology are allowed in fibration, whereas topologically different leaves (the analogue of fibers) are allowed in foliation.
Formally, a foliation atlas of codimension $q$ of $M$ 
is a collection of the coordinate patches
%%%
\bea
\vf_i: U_i \ra R^n=(R^{n-q}\times R^{q})
\eea
%%%
The local coordinate transformation $\vf_{ij}$ between $\vf_i$'s takes the form of
%%%
\bea
\vf_{ij}(x,y)=(\vf_{\1 ij}(x,y),\vf_{\2 ij}(x,y))
\eea
%%%
where $\vf_{\1 ij}(x,y)$ (resp. $\vf_{\2 ij}(x,y)$) is associated with $R^{n-q}$ (resp. $R^{q}$).  
 The $(n-q)$-dimensional submanifold (injectively immersed) in $M$ is called a leaf.
When the codimension is $q=1$, the case of our focus, the leaves are called hypersurfaces.

A globally hyperbolic spacetime that we focus on in this work admits foliation through a family of hypersurfaces $\Sigma_t$; the base manifold is parameterized by a ``time" coordinate $t$.  
Let us choose a coordinate system such that a vector $\bfX$ takes
%%%
\bea
\bfX\equiv {\bf\pa}_\a =(\pa_t,\pa_a),\quad a=1,2,3
\eea
%%%
and resolve $\pa_t$ according to
%%%
\bea
\pa_t=n\hat{\bfn}+N^a \pa_a
\eea
%%%
where $\pa_a$ is a vector tangent to $\Sigma_t$; $n$ is the lapse function and $N^a$ is the shift vector.
In the present coordinate system, the components of the metric tensor $g_{\a\b}\equiv \bfg(\pa_\a, \pa_\b)$ are given in the conventional notation by
%%%
\bea
ds^2=g_{\a\b} dx^\a dx^\b=(-n^2 +h^{ab}N_aN_b)dt^2+N_a dtdy^a+h_{ab}dy^a dy^b
\eea
%%%
The foliation has 3D leaves whose space is described by $h_{ab}$. The space of the leaves (i.e., the base manifold) is parameterized by $t$ (whose component is denoted by $t^\a$).

As stated in the introduction, we will see that the condition obtained in \cite{Park:2014tia} by examining the shift vector constraint can be related to the condition for the foliation to be Riemannian. Then in the ``dual" view explained in section \ref{main}, the manifold admits the so-called totally geodesic foliation whose definition is given in terms of the second fundamental form.  
The second fundamental form $\bfK$, which is also called the extrinsic curvature, of a given hypersurface $\Sigma_t$ is defined by
%%%
\bea
\bfK(\bfA,\bfB)=-\bfg(\bfA,\N_\bfB\, \hat{\bfn})
\eea
%%%
where $\hat{\bfn}$ is the unit vector normal to the hypersurface and $\bfA, \bfB$ represent the vectors tangent to the hypersurface.
When $\bfK$ vanishes, the foliation is called totally geodesic.

%%%%%%%%%%%%%%%%%%%%%%%%%%%%%%%%%%%%%
%%%%%%%%%%%%%%%%%%%%%%%%%%%%%%%%%%%%%
\section{ Quantization via hypersurface}
%%%%%%%%%%%%%%%%%%%%%%%%%%%%%%%%%%%%%
%%%%%%%%%%%%%%%%%%%%%%%%%%%%%%%%%%%%%
With the review on differential geometry in the previous section, we are ready to approach the gravity holography as observed in \cite{Park:2014tia} from a more mathematical perspective. Let us briefly review the findings in \cite{Park:2014tia} before we get to the main analysis in subsection \ref{main}.

\subsection{Review of gravity holography}
Consider the 4D Einstein-Hilbert action 
%%%
\bea
S=\int d^4 x \sqrt{-g}\;R  \la{unsplit}
\eea
%%%
and the operator quantization.
We split the coordinates into
%%%
\bea
x^\m\equiv (t,y^a)
\eea
%%%
where $\m=0,..,3$ and $a=1,2,3$.
By parameterizing the 4D metric \cite{Arnowitt:1962hi}\cite{Poisson} in the the 1+3 split form 
%%%
\bea
g_{\m\n}=\left(
\begin{array}{cc}
-n^2+h^{ab}N_{a} N_{ b} & N_{ a} \\
&\\
N_{ b} & h_{ab}
\end{array}
\right) 
\eea
%%%
where $n$ and $N_{ a}$ denote the lapse function and shift vector respectively,
one gets
%%%
\bea
S=\int d^4 x\;n\sqrt{-h} \left(R^{(3)}-K^2+K_{ab}K^{ab}\right)
\la{1p3act}
\eea
%%%
with the second fundamental form given by
%%%
\be
K_{ab}=\fr1{2n}\left(\mathscr{L}_t h_{ab}-{\nabla}_a N_{b}
         -{\nabla}_b N_{ a} \right),\qquad K=h^{ab}K_{ab}.
\la{K4defqq}
\ee
%%%
Here $\mathscr{L}_t$ denotes the Lie derivative along the time coordinate $t$ and $\N_a$ is the 3D covariant derivative with the connection form constructed out of the hypersurface metric $h_{ab}$. Since the time derivative does not act on $N_a$ or $n$ in their field equations
%%%
\bea
{\N}_a (K^{ab}-h^{ab} K)=0  \la{Ncon}
\eea
%%%
\bea
R^{(3)}+K^2-K_{ab}K^{ab}=0  \la{ncon}
\eea
%%% 
these fields are non-dynamical: once $n$ and $N_{ a}$ are specified on the hypersurface of a given time, their bulk value can be taken as the corresponding value on the hypersurface of the fixed time.
The 4D diffeomorphism can be fixed by imposing the de Donder gauge. The action \rf{1p3act} still has 3D gauge symmetry of measure-zero compared with the 4D gauge symmetry. By using this 3D diffeomorphism, the shift vector can be gauged away
%%%
\bea
N_{a}=0,   \la{Nazero}
\eea
%%%
in the entire bulk due to the non-dynamism of $N_a$.
Substituting $N_{ a}=0$ into \rf{Ncon} it follows that
%%%
\bea
{\N}^a \left[\fr{1}{n}\Big(\mathscr{L}_t h_{ab}
 -h_{ab}h^{cd}\mathscr{L}_t h_{cd}\Big)\right]=0  \la{mtmconstr}
\eea
%%%
which implies
 %%%
 \bea
 \pa_a n=0 \la{constronn}
 \eea
 %%% 
 This can be seen as follows. What we need to show is that 
 the covariant derivative in \rf{mtmconstr} yields zero when it acts on terms 
 other than $\fr1{n}$. Let us illustrate this with the first term in the parenthesis:  
%%%
\bea
 \N_a \mathscr{L}_t h_{bc}=e_a^\a  \N_\a \mathscr{L}_t h_{bc}
\eea
%%% 
By using \rf{liecocom}, the right-hand side can be written as
%%%
\bea
 =e_a^\a   \mathscr{L}_t \N_\a h_{bc}
\eea
%%%
This is because $\N_{[\bfX,\bfY]}=\N_{[\pa_t,\pa_\a]}=\N_{0}=0$ due to the linearity of $\N$. On account of $\mathscr{L}_t e_a^\a=0 $, the right-hand side becomes
%%%
\bea
= \mathscr{L}_t e_a^\a  \N_\a h_{bc} =\mathscr{L}_t   \N_a h_{bc}=0
\eea
%%%
where the last equality follows from the 3D metric compatibility of the 3D covariant derivative. In the next subsection, we will note that the condition \rf{constronn} is nothing but the requirement for the codimension-1 foliation to be Riemannian.

\subsection{TG foliation and gauge symmetry \la{main}}
In this subsection, we make several crucial observations that lead to holographic reduction of the bulk to the hypersurface. Firstly, we relate \rf{constronn} to the condition for the foliation to be Riemannian. Afterwards, a dual view of the totally geodesic foliation of codimension-3 is taken. It was proven in \cite{Molino}\cite{Cairns} that a totally geodesic foliation carries Lie algebra.\footnote{More precisely speaking, the pullback of the geodesic to the frame bundle develops a so-called "tangential parallelism" which then leads to a Lie algebra that acts transitively on the leaves of the pullback bundle.}  As our last crucial step, we identify the Lie algebra with diffeomorphism pertaining to the 4D metric. We will elaborate on this identification below, but let us first relate \rf{constronn} to Riemannian foliation.

The condition \rf{constronn} obtained by the shift vector constraint can be written as
%%%
\bea
\mathscr{L}_{\pa_a} n=0  \la{Riemannconp}
\eea
%%%
This is precisely the condition for the foliation to be Riemannian.
To see this, let us denote the horizontal component of the metric tensor $\bfg$ by $\bfg^h$:
%%%
\bea
\bfg^h(X,Y)\equiv \bfg(X^h,Y^h)
\eea
%%%
The Riemannian foliation satisfies  
%%%
\bea
\mathscr{L}_{X} \,\bfg^h=0,\quad \bfX\in {\cal V}  \la{Riemanncon}
\eea
%%%
by definition (see, e.g., \cite{Gromoll}). The lapse function $n$ corresponds to the $(t,t)$ component of $\bfg^h$.\footnote{The manifold is viewed as having 3D leaves and 1D space of leaves. Recall that the $\pa_a$-directions must be tangent to the leaves.}
In the ``dual" view, the Riemannian foliation implies totally geodesic foliation (see, e.g.,\cite{Dotto}). In other words, the manifold can be viewed as 1D fibration over the 3D base, and {the 1D fibration will be totally geodesic.} Then there should be an abelian Lie group associated with the fibration \cite{Cairns}. This should be the abelian group that acts along the 1D fiber in the dual picture. This in turn should imply that the manifold can be constructed as U(1) fibration over the 3D base.

To rephrase, let us now take the dual view wherein the original Riemannian foliation of codimension-1 is viewed as the totally geodesic foliation of codimension-3.
It was proven in \cite{Cairns} that a totally geodesic foliation carries Lie algebra\footnote{More precisely speaking, the pullback of the geodesic to the frame bundle develops a so-called "tangential parallelism" which then leads to a Lie algebra that acts transitively on the leaves of the pullback bundle.}; in the present case, the Lie algebra is abelian. As our last important step, we identify the Lie algebra with diffeomorphisms pertaining to the 4D metric.
The lapse function and shift vector concern displacements away from the hypersurface whereas the induced hypersurface metric concerns displacements within the hypersurface (see, e.g., \cite{Poisson}). This naturally seems to suggest that the gauge symmetry be associated with the action of group fibration that generates the 4th direction. The gauge-fixing then corresponds to taking the quotient of the bundle by the group, bringing us to the holographic reduction of the physical states.\footnote{Once the external states of the Feynman diagrams are restricted to these physical states, the renormalizability is achieved \cite{Park:2014tia,Park:2016zgt}. In the conventional approach, the well-known offshell non-renormalizability was established in the seventies: the renormalizability achieved in \cite{Park:2014tia,Park:2016zgt} pertains to the physical states defined by the lapse function and shift vector constraints.}

%%%%%%%%%%%%%%%%%%%%%%%%%%%%%%%%%%%%%
%%%%%%%%%%%%%%%%%%%%%%%%%%%%%%%%%%%%%
\section{Conclusion}
%%%%%%%%%%%%%%%%%%%%%%%%%%%%%%%%%%%%%
%%%%%%%%%%%%%%%%%%%%%%%%%%%%%%%%%%%%%
In this work, we have related the shift vector constraint to the requirement for the foliation to be Riemannian. Then in the dual picture, the foliation by 1D leaves is totally geodesic. A totally geodesic foliation has an associated Lie algebra as proven in \cite{Cairns}. In our case, it is abelian and we have identified it as the origin of the diffeomorphism of the 4D metric.

In a complementary view, one may say the following. Fibering generates one more dimension, i.e, 3D becomes 4D. On top of the 3D metric, one now has four more metric components that correspond to shift and lapse. Since the physics is governed by the action - which has gauge invariance - but not directly by the metric, not all of those new metric components would be physical. As a matter of fact, none of them is physical.  
Therefore, the generation of the 4th direction should be associated with the diffeomorphism. Based on this, we have proposed that the abelian fibration at the level of spacetime be associated with the diffeomorphism at the level of the metric configuration bundle.

It is presumably the causality property of a globally hyperbolic spacetime that is responsible, on a deeper level, for the reduction of the bulk degrees of freedom.
It would be interesting to make this precise. 
The mathematics used in this work is rather abstract. It would be worthwhile to expand the contents of this work in a more self-contained and physicist-accessible form. Some of the mathematical results and their physical meanings should also be explored more thoroughly. We also plan on explicitly carrying out the procedure of quantizing through the hypersurface in the near future. It is expected that the renormalization procedure will be technically demanding, and it will require care to establish the precise way of extracting 4D physics through the hypersurface physics. We will report on these tasks elsewhere.

\medskip
\ni {\bf Note added.} It was a pleasant surprise to find out about the works of \cite{York:1972pr}\cite{Isenberg}\cite{Moncrief:1989dx}\cite{Fischer:1996qg} a few months after the completion of this work. In some of those works,  
the authors used the Hamiltonian approach to show that gauge-fixing reduces the 4D Hamiltonian into the 3D Hamiltonian. A more recent related discussion can be found in \cite{Gerhardt:2012ku} with which the present manuscript has a certain overlap in spirit. The new ingredients of the present work are the role played by the totally geodesic foliation and the proposal that the abelian fibration at the level of spacetime be associated with the diffeomorphism at the level of the metric configuration bundle. Various ideas in the present work have been further developed; a comprehensive list of the subsequent works can be found in \cite{Park:2018vci}.

%\vspace{.3in}

%\ni {\bf Acknowledgments}

%\ni The research of this work was funded in part by Hangyang University, South Korea.

\newpage
%%%%%%%%%%%%%%%%%%%%%%%%%%%%%%%%%%%%%%%%%%%%%%%%%%%%%%%%%%%%%%%%

\end{document}